\documentclass[a4paper]{article}

\usepackage{INTERSPEECH2021}

\usepackage{amssymb}
\usepackage{multirow}
\usepackage{subdefs}
\usepackage{multicol}
\usepackage{caption}
\usepackage{subcaption}
\usepackage[ruled,vlined]{algorithm2e}

\usepackage{hyperref}

\title{Cross-Modal learning for Audio-Visual Video Parsing}
\name{Jatin Lamba, Abhishek, Jayaprakash Akula, Rishabh Dabral, Preethi Jyothi, Ganesh Ramakrishnan}

\address{
  Indian Institute of Techonology, Bombay}
\email{\{jatinl, abhishekthakur, jayaprakash,
rdabral, pjyothi, ganesh\}@cse.iitb.ac.in}

\begin{document}

\maketitle

\begin{abstract}
In this paper, we present a novel approach to the audio-visual video parsing (AVVP) task that demarcates events from a video separately for audio and visual modalities. The proposed parsing approach simultaneously detects the  temporal boundaries in terms of  start and end times of such events. We show how AVVP can benefit from the following techniques geared towards effective cross-modal learning: (i) adversarial training and skip connections (ii) global context aware attention and, (iii) self-supervised pretraining using an audio-video grounding objective to obtain cross-modal audio-video representations. We present extensive experimental evaluations on the Look, Listen, and Parse (LLP) dataset and show that we outperform the state-of-the-art Hybrid Attention Network (HAN) on all five metrics proposed for AVVP. We also present several ablations to validate the effect of pretraining, global attention and adversarial training.  \footnote{Code and other details are  available at \url{https://www.cse.iitb.ac.in/~malta/avvp}}
\end{abstract}

\maketitle
\noindent\textbf{Index Terms}: audio-visual video parsing, cross-modal learning. 

\section{Introduction}
Audio Visual Video Parsing (AVVP)~\cite{tian2020avvp} is a newly introduced multi-modal task that involves detecting and localizing occurrences of events within the audio and visual streams of a video. AVVP has numerous potential applications. It can directly contribute to audio-visual source separation~\cite{Ephrat_2018}, especially when the sources of audio are occluded in the video. AVVP could also feed into downstream video understanding tasks (such as captioning or summarization) that could benefit from both audio and visual cues.  

AVVP has been formulated as a Multi-modal Multi-Instance Learning (MMIL) task. The task becomes particularly challenging when there is only weak supervision; the set of events occurring in the audio and visual streams are available as a bag of events, while the start and end times of each individual event are not available. It is also entirely possible to have asynchronous annotations with different start and end times for the same event in different modalities. Furthermore, it is common to observe certain events (such as human speaking, telephone ringing, singing, baby crying, dog barking, violin playing, car, and vacuum cleaning) occurring either only in the audio modality or only in the visual modality. 

\begin{figure}
\includegraphics[width=0.5\textwidth]{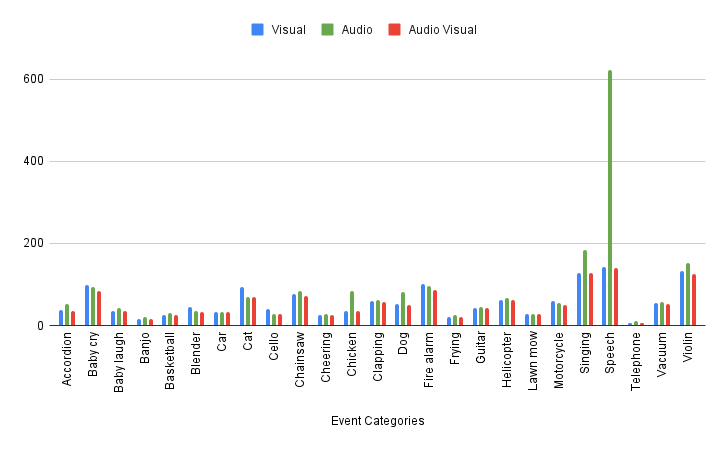}
\caption{Distribution of audio, visual and audio-visual events in the LLP dataset~\cite{tian2020avvp}, across the $25$ event categories. Note that the three lines blue, green and red are not disjoint sets.}
\label{fig:stats}
\end{figure}

Existing methods~\cite{tian2020avvp} have attempted to solve this task by learning cross-modal audio-visual features wherein representations for each modality are expected to benefit from the contexts surrounding the other modality. While such cross-modal representations are indeed powerful, they may not always be beneficial across modalities and might come at the cost of event detection performance for a specific modality. 

For the Look, Listen Parse (LLP) dataset~\cite{tian2020avvp}, we believe the audio modality may not always benefit from the proposed cross-modal architecture for the following reasons. \textit{Firstly,} there exists a dataset skew in favour of the audio modality that renders video sequences inconsequential for detecting audio streams. E.g., there are approximately 600 videos with speech as an audio event, while only 120 videos list it as a video event. In Figure~\ref{fig:stats} we present the distribution of audio, visual and audio-visual events across all event categories. As another example, visual features may be useless if the \textit{dog bark} audio event occurs in the background of the scene. \textit{Secondly}, it may be argued that input features for audio (VGGish) \cite{vgg} are already well suited for audio parsing and the inclusion of cross-modal information from visual stream introduces noise into the resulting features. \textit{Finally}, the existing state-of-the-art architecture for AVVP~\cite{tian2020avvp} attempts to project audio and visual features to a shared event label space using a common fully connected layer. Such a common projection layer might hurt certain audio events that do not benefit from visual features. For example, \textit{doorbell ringing} might often be an audio event in the background without any supporting visual cues.

To alleviate these issues, we attempt to improve the performance of both visual and audio modalities. To that end, we introduce an adversarial loss to improve the quality of the shared audio-visual features. We alter the architecture to ensure that the audio features are not always influenced by cross-modal attention from the visual stream. We also leverage a global context attention mechanism \cite{gcaa} so that event prediction benefits not only from temporally local features but also from the global video context. Finally, we explore the benefits of using pretrained representations trained using self-supervised objectives (such as audio-video grounding) on a large audio dataset.

With our proposed techniques, we observe significant improvements not only in audio event detection performance but also in video detection performance on the Look, Listen and Parse (LLP) dataset~\cite{tian2020avvp}. Specifically, we improve the audio, visual and  audio-visual F1-scores from 60.1\% to 61.6\%, 52.9\% to 54.7\% and 48.9\% to 50.3\%, respectively. 

\section{Related Work}

\subsection{Cross-Modal Attention Networks} 
Most of the work done in audio-visual learning assumes that temporal synchronized audio and visual content convey the same semantic meaning. However, natural unconstrained videos are noisy and tend to have redundant audio visual events that recur many times in the video, both within the same modality~\cite{1039875} as well as across different modalities~\cite{NEURIPS2018_c4616f5a},\cite{1a18f054b93d4818af72164f56836616}. Hybrid Attention Network (HAN) \cite{tian2020avvp} tries to jointly model the modalities and any asynchrony between them in a unified manner via self-attention and cross-attention networks. Xuan {\em et. al.}~\cite{Xuan_Zhang_Chen_Yang_Yan_2020} propose a cross-modal attention network for audio-visual event localization with modality-specific CNNs acting on each modality to produce feature maps. These are subsequently fed into spatial, global context-aware and cross-modal attention modules that respectively learn `where', `when' and `which' modality to attend to for event localization. We adopt the idea of a global context-aware attention mechanism which reflects the contribution of each individual segment towards event localization. Audio-visual speech recognition has also benefitted from cross-modal architectures~\cite{dcma}. In this work, a dual cross-modality attention scheme is proposed for a transformer-based model that combines two cross-modality attentions. DCM attention magnifies the role of visual modality to the same level as that of audio modality thus yielding better performance. 

\subsection{Multiple Instance Learning (MIL)} 
Multiple instance learning (MIL)~\cite{mil} is a form of weakly supervised learning, wherein training instances are arranged in sets, called bags, and a label is provided for the entire bag instead of each element of the bag. Using a collection of labeled bags, the learner tries to induce a concept that will either (i)  label individual instances correctly or (ii) label bags. Unlike the previous audio-visual event localization  works~(e.g., \cite{tian2018audiovisual}) that are formulated as an MIL problem where an audio-visual snippet pair is regarded as an instance, each audio snippet and the corresponding visual snippet occurred at the same time denote two individual instances in a bag in our task. The multimodal multiple instance learning (MMIL) method in AVVP respects audio-visual temporal asynchrony.

\subsection{Self-supervised learning}
There have been previous attempts toward self-supervised audio-visual pre-training such as ~\cite{audiovisualscene,avid,avcluster}. In~\cite{audiovisualscene}, in a self supervised manner, the authors learn a temporal, multisensory representation  that fuses the visual and audio components of a video signal. They train a neural network on a pretraining task of detecting misalignment between audio and visual streams in synthetically-shifted videos. In order to detect misalignment in a video of human speech, the integration of low level information across modalities is required, for which they propose a 3D multisensory convolutional network (CNN) with early fusion of audio and visual streams for modeling actions that produce a signal in both modalities.

\section{Our Approach}
\label{sec:method}

AVVP requires temporally grouping (contiguous) video snippets into  audio, visual, and audio-visual events,  and associating semantic labels with each event. %
More specifically, let $\Scal$ be a set of event categories. In the case of LLP, $\Scal$ has $25$ events categories; {\em e.g.}, {\em human speaking}, {\em singing}. %
Given a video sequence containing both audio and visual tracks, the sequence is divided into $T$ non-overlapping audio and visual snippet pairs $\{ V_t, A_t \}_{t=1}^{T}$, where each snippet is 1 second long. $V_t$ and $A_t$ respectively denote the visual and audio content within the video snippet.
Let 
\begin{equation}
 y_t = \left\{(y^a_t, y^v_t, y^{av}_t ) \left|\ {[y^a_t]}_s, {[y^v_t]}_s, {[y^{av}_t]}_s \in \{0, 1\},s \in \Scal \right. \right\} 
\end{equation} be the event label
set for the video snippet $\{V_t, A_t\}$, where $s \in \Scal$ refers to an event category and $y_t^a$, $y_t^v$, and $y_t^{av}$ denote audio, visual, and audio-visual event labels, respectively. 
Here, we have a relation: $y_t^{av}= y_t^a*y_t^v$, which means that audio-visual events occur only when there exist both audio and visual events during the same time span and and
from the same event categories.
\begin{figure*}
\centering
\includegraphics[width=0.8\textwidth]{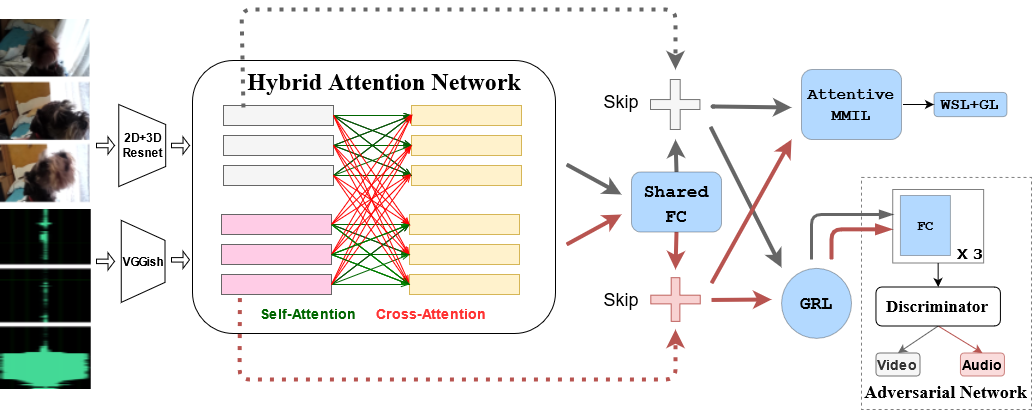}
\caption{Depiction of architectural changes made to the HAN network~\cite{tian2020avvp}. The blocks in the cyan colour highlight modifications to the adversarial network.}
\label{fig:han} %
\end{figure*}
We leverage the state-of-the-art Hybrid Attention Network (HAN)~\cite{tian2020avvp} as the base for our network architecture. 
Given separate streams of pre-trained audio (VGGish~\cite{vgg}) and visual (ResNet-152~\cite{resnet152} and 3DResNet~\cite{r2p1d} features), HAN uses cross-modal attention to learn modality contextualized features. 
As described in Figure~\ref{fig:han},  these features are then provided as an input to a shared linear layer that produces joint features for audio and video before finally passing them for Attentive MMIL Pooling. 

\noindent \textbf{Dataset Bias in LLP Dataset:}
We observe in Table 2 of ~\cite{tian2020avvp} that improvement of visual event parsing degrades the performance of audio event parsing. This may be attributed to the fact that the LLP dataset ({\em c.f.} Figure~\ref{fig:stats} for stats) is itself collected from an audio oriented dataset, {\em viz.}, AudioSet. AudioSet has 1447 events in the video domain, as against a much higher 2090 events in the audio domain. %
Overall, this leads us to hypothesize that learning joint features through cross-modal attention and a shared linear layer, though beneficial for visual features, could be detrimental to the quality of audio features.

\subsection{Our Proposed Model}
\label{sec:model}
In order to address the aforementioned issue of bias in the LLP dataset~\cite{tian2020avvp} affecting the AVVP task, we propose to (a) use the joint features as residuals to the self-attention outputs through skip connections, (b) introduce an adversarial loss to improve the learnt joint-features and (c) use global context attention for learning higher level features. See Figure~\ref{fig:han} for our architecture.
\vspace{0.2em}

\begin{figure}[th!]
\includegraphics[width=0.5\textwidth]{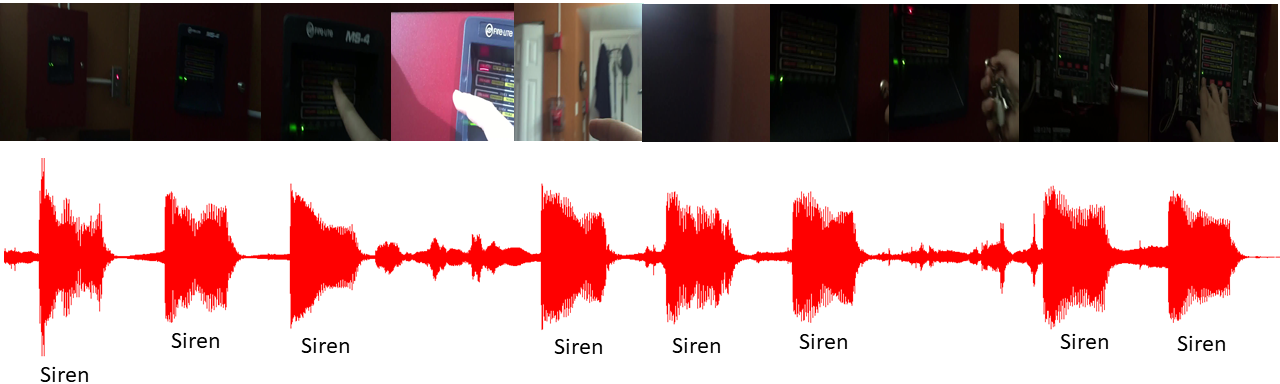}
\caption{The video and audio correspond to a fire alarm event. The video frames have no cues relevant to fire alarm. But only from the audio using the \texttt{siren} sound, the model can understand the actual event}
\label{fig:siren}
\end{figure}

\noindent \textbf{Skip connections:}
Skip connections are used in the attention module to pass self attended signals over the cross attention module. Using skip connections via addition, we try to preserve  pure signals from each modality to avoid cluttering of unnecessary information from the other modality. Since LLP is rich in audio data (again see Figure~\ref{fig:stats}), audio representation should not require as much support from visual data ({\em c.f.}, Figure~\ref{fig:siren}) as the support visual representation might require from audio. The effectiveness of skip connections on audio is evident from Table~\ref{tab:avsegment}. 
\vspace{0.2em}

\noindent \textbf{Adversarial Training:}
The attention network tries to learn information across both modalities which can be essential for the audio visual parsing task. In order to improve upon learning of joint features across modalities, just prior to the MMIL pooling layer, we add a modality discriminator to learn to discriminate between audio and visual features. %
During training, the parameters of the underlying attention network are optimized in order to minimize the loss of the video parser classifier and to maximize the binary cross entropy loss of the modality discriminator. In practice, a gradient reversal layer~\cite{gradientReversal16} is added over the features which feeds into the modality discriminator. Concretely, for a feature vector $f \in \mathbb{R}^d$, where d is the hidden dimension, let $G_m(f, \theta_m)$ be the modality discriminator with parameters $\theta_m$. With the ground truth modality label
$y_m$, we optimize the proposed discriminator with a binary cross-entropy function: 
\begin{equation}
    \mathcal{L}_{ad}=-(y_m\log(G_m(f, .) + (1-y_m)\log(G_m(f, .))
    \label{sec:advloss}
\end{equation} 

\noindent \textbf{Global Context Aware Attention:}
We replace the multi-head attention in HAN with global context aware attention~\cite{gcaa}. A global context vector is computed, taking average over the features of   the video segments to represent the global meaning of the entire sequence. The local context of a segment is modified using the segment level representation and the global context vector. These local context vectors are employed to obtain attention similar to the scaled dot product attention in transformers. We use this modified attention module for both the self and cross modal attention networks. 

\vspace{0.2em}
\noindent \textbf{Overall Loss Objective:} Our  overall loss can be expressed as:
\begin{equation}
    \mathcal{L} = \mathcal{L}_{wsl} + \lambda_g\Lcal_g - \lambda_{ad}\Lcal_{ad}
    \label{sec:overallloss}
\end{equation} 
where $\mathcal{L}_{wsl}$ and $\mathcal{L}_{g}$ are borrowed from~\cite{tian2020avvp} and $\mathcal{L}_{ad}$ is the adversarial loss discussed above. $\Lcal_{wsl}$ is the main weakly supervised loss~\cite{tian2020avvp} for event classification while $\Lcal_g$ is the modality specific classification loss that is applied after label smoothing. 

\subsection{Pretraining for Improved Features}
\label{sec:pretraining}

In the context of our model proposed in Section~\ref{sec:model}, we investigate if pretraining the audio and visual input features could further enhance the performance of our  model on the AVVP task.
To this end, we introduce a novel pretraining strategy that jointly trains audio and visual inputs using the task of aligning the audio and visual modalities.  \\

\indent For pretraining through audio-visual alignment, we adapt the architecture of UNITER \cite{uniter} and accommodate the modifications necessary for the audio and visual modalities. UNITER is a large scale image-text embedding network. It adopts the transformer architecture as the core and leverages its self attention mechanism to learn representations for both modalities in a joint embedding space. These joint representations are learnt by simultaneous application of four different self-supervised pre-training objectives {\em viz.}, Masked Language Modeling (MLM), Masked Region Modeling (MRM), Image-Text Matching (ITM), and Word-Region Alignment (WRA).

We use the pretrained embeddings for audio and visual modalities extracted from VGGish and Resnet152 networks. The pretrained and temporal features for both the modalities are then passed through a fully-connected (FC) layer, to obtain the embedding vectors. These embeddings are then
fed into a multi-layer Transformer to learn a cross-modal contextualized embedding
across video and audio clips.
Note, that audio-visual alignment can be treated as a problem in itself and here we use the alignment task purely for the purpose of pretraining.
\begin{figure*}
\centering
\begin{subfigure}{.45\textwidth}
\centering
\includegraphics[width=\linewidth]{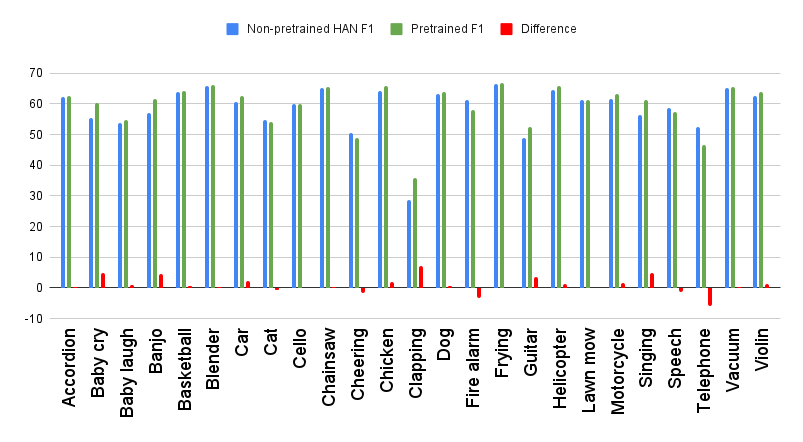}
    \label{fig:aud_res}
\end{subfigure}
\begin{subfigure}{.45\textwidth}
    \centering
    \includegraphics[width=\linewidth]{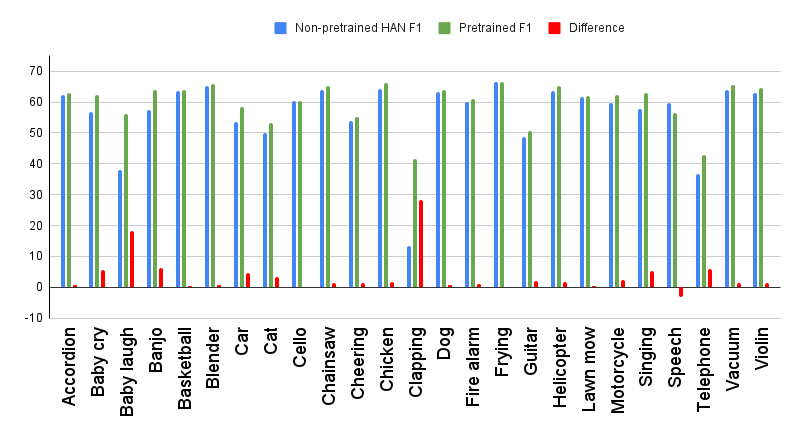}
    \label{fig:vid_res}
\end{subfigure}
\caption{Category-wise F1 scores from the non pretrained HAN and pretrained  models for the audio (left) and visual (right) modalities.}\label{fig:av_res}
\end{figure*}

\textbf{Audio-Visual Grounding (AVG) as a Pretraining Task:} The inputs to AVG are a video clip and an audio clip sampled at random from a video segment and an audio segment,  respectively. A concatenated audio-visual embedding is fed into our model as a fused representation of both modalities. These embeddings are then trained to be discriminative  based on the contrastive loss.

We extract the joint representation of the input snippet pairs, using cosine similarity, to  predict a score between 0 and 1. 
AVG supervision is over the predicted binary labels. During training, we sample an audio-visual snippet pair $\langle a_i, v_i \rangle$ and ground it with respect to all video-audio snippets. We apply the  contrastive loss $\mathcal{L}$ as in \eqref{eq:uni-avg}\eqref{eq:cross-avg} for pretraining, where $p$ and $n$ are the margin parameters for the positive and negative pairs respectively. ($[]_+$ refers to max(.,0).) 
We try several variants of AVG-based objectives: (i) Uni-modal $\text{AVG}_{\text{u-avg}}$, (ii) Cross-modal $\text{AVG}_{\text{x-avg}}$ and (iii) Multi-modal grounding $\text{AVG}_{\text{m-avg}}$ shown in equations~\eqref{eq:uni-avg}, \eqref{eq:cross-avg} and~\eqref{eq:multi-avg}, respectively. We provide ablations in Section~\ref{sec:expts} demonstrating the effectiveness of these variants.
\begin{equation}
\begin{aligned}
	\mathcal{L}_{\text{u-avg}} =  \sum_{j}^{i,j \in Pos}[p - \langle a_i,a_j\rangle ]_+ +  \sum_{j}^{i,j \in Neg}[\langle a_i,a_j\rangle - n]_+ \\ +\sum_{j}^{i,j \in Pos}[p - \langle v_i,v_j\rangle ]_+ +  \sum_{j}^{i,j \in Neg}[\langle v_i,v_j\rangle - n]_+
	\label{eq:uni-avg}
\end{aligned}
\end{equation}
\begin{equation}
\begin{aligned}
	\mathcal{L}_{\text{x-avg}} =  \sum_{j}^{i,j \in Pos}[p - \langle a_i,v_j\rangle ]_+ +  \sum_{j}^{i,j \in Neg}[\langle a_i,v_j\rangle - n]_+ \\ +\sum_{j}^{i,j \in Pos}[p - \langle v_i,a_j\rangle ]_+ +  \sum_{j}^{i,j \in Neg}[\langle v_i,a_j\rangle - n]_+
	\label{eq:cross-avg}
\end{aligned}
\end{equation}
\begin{equation}
	\mathcal{L}_{\text{{m-avg}}} = \mathcal{L}_{\text{{u-avg}}} + \mathcal{L}_{\text{{x-avg}}}
	\label{eq:multi-avg}
\end{equation}
where is i is any random segment from the audio-visual clip.

\textbf{Obtaining ground truth for AVG}
The binary ground truth for each audio-video snippet pair could have been based on exact match of their temporal spans. However an audio snippet could also loosely (semantically) correspond to several video snippets with   spans different from its own. Hence, a one-to-one mapping  based on temporal span match might itself not be effective. Hence, we generate the ground truth using the method described next. Instead of simple clustering based grouding where-in the snippets  are clustered in their representation space, we take a graph based connectivity approach for obtaining the ground truth as described below.

\begin{figure}[h!]
\centering
\includegraphics[width=0.45\textwidth]{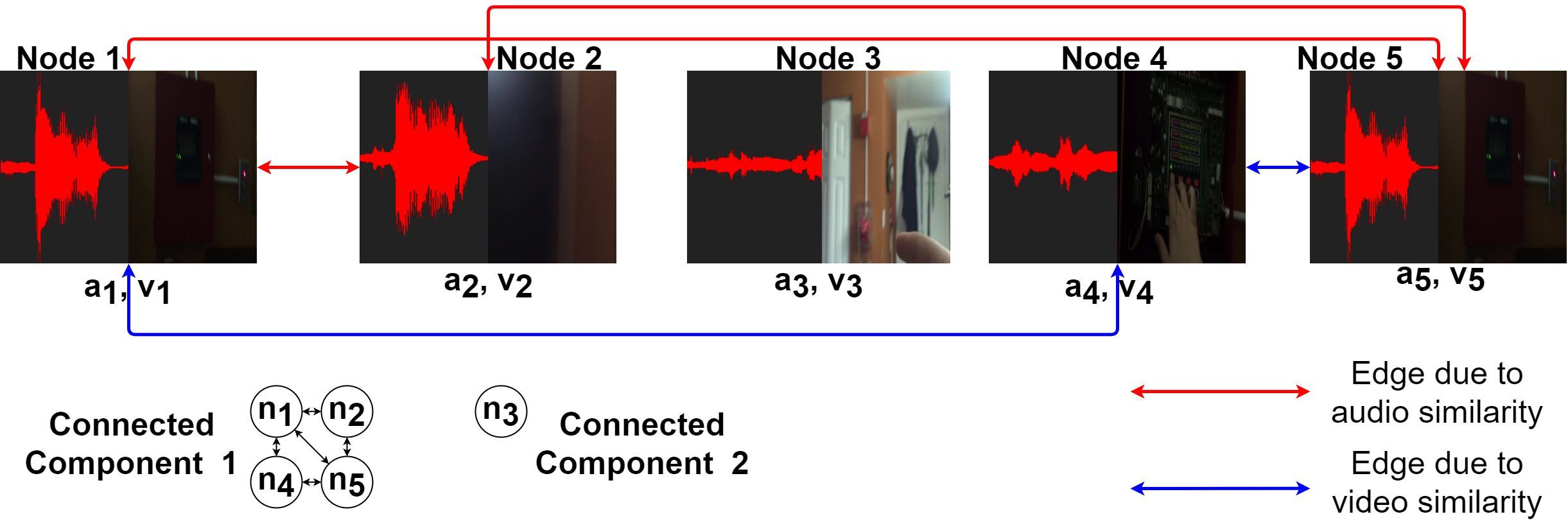}
\caption{A detailed illustrative description of the procedure to obtain the grounding labels. Node 1, 2, 4 and 5 belong to same group while node 3 is in different group.}\label{fig:connected} %
\end{figure}

We construct a graph where each snippet is represented by a node and nodes are connected only if the corresponding snippets  are {\em semantically similar}. 
Further, we consider two snippets  to be {\em semantically similar} if either (i) the Resnet152 embeddings of corresponding video snippet or (ii) VGGish representation  for the corresponding audio snippet, have similarity above some threshold. This is a adaptation of agreement score as in \cite{avid}, where similarity score used to determine the positives and negatives required in their setup.
Now, all the audio-video snippets across different  pairs (nodes) in the same connected component in this graph are considered to have ground truth label 1. Any other pair of audio-video snippets is considered to have ground truth label 0 see figure \ref{fig:connected}.

\begin{algorithm}
\KwIn{$(V, A)$ set of video-audio dataset}
\KwOut{The ground truth for grounding task}
\KwData{$S_i=(V_i,A_i)$ such that $V_i, A_i$ are the i-th video-audio pair}
$s_{ij} = (v_{ij}, a_{ij}), j \in \{1,2,\dots, 10\}$ ie 1-sec long video-audio snippet pair embeddings.\\
\textbf{initialisation}: Graph $G_i$ for the i-th video-audio pair\\
A node $n_{ij}$ for each video-audio snippet pair.\\
\For{each pairs of snippets $(s_u, s_v)$ in $S_i$}{
$GT(u, v) \gets 0$
}
\For{$j \gets 1$ \textbf{to} $10$}{
\For{$k \gets j+1$ \textbf{to} $10$}{
\If{$\langle v_{ij}, v_{ik} \rangle \geq v_{t} $ \textbf{or} $\langle a_{ij}, a_{ik} \rangle \geq a_{t} $} {
    connect node $n_{ij}$ and $n_{ik}$;
 }
 }
 }
Obtain the connected components $C_i$in $G_i$. \\
\For{\text{each component }$C$\text{ in }$C_i$}{
\For{\text{each pairs of nodes }$(u, v)$\text{ in }$C$}{
$GT(s_u, s_v) \gets 1$
}
}
\caption{Algorithm describing the procedure to obtain the ground truth for AVG task}
\end{algorithm}

\section{Experiments and Results}
\label{sec:expts}

\noindent \textbf{Experimental Setup:}
The LLP dataset consists of videos each of duration 10 secs. We divide each video into ten segments, each 1 sec long. For a visual segment, ResNet152~\cite{resnet152} features are extracted over frames sampled at 8fps and later fused with 3D ResNet to obtain 512-dimensional segment-level visual features. For an audio segment, we use the VGGish extractor~\cite{vgg} to get  512-dimensional audio features. We set $\lambda_g=0.6$ and $\lambda_{ad}=0.4$ in our objective function in Eqn.\eqref{sec:overallloss}. We train our model with a learning rate of 3e-4 and decay it by a multiplicative factor of 0.5 every 5 epochs. We train our models for 40 epochs with a batch size of 64.
We pretrain using the audio visual grounding objectives ({\em c.f.}, Section~\ref{sec:pretraining}), on a subset (29K videos) of AudioSet~\cite{audioset} and extract representations for both audio and visual streams. 

\vspace{0.5 em}
\noindent \textbf{Results and Analysis:} We report results using the same evaluation metrics as in~\cite{tian2020avvp} (and elaborated in the extended version). Tables~\ref{tab:avsegment} and~\ref{tab:avevent} present evaluations at the segment and event levels, respectively. `Audio', `Visual' and `AV' refer respectively to audio, visual, and audio-visual events. Segment-level F-scores are evaluated at the level of segments, while the event-level F-scores are computed by concatenating  consecutive positive snippets in the same event categories and then computing an event-level F-score based on mIoU = 0.5 as the threshold. We also compute two aggregate metrics, {\em viz.}, Type@AV (Macro F1) and Event@AV (Micro F1) which are averaged audio, visual, and audio-visual event evaluation results and the F-scores considering all events for each sample, respectively.

HAN is the current state-of-the-art system~\cite{tian2020avvp} for AVVP. `Base' differs from HAN in: (i) the use of separate self attention modules for the two modalities (HAN uses a shared module) and (ii) the use of modality fusion over self-attended features (HAN performs cross modal attention directly over encoded features). These modifications to  `Base' were beneficial when used in conjunction with our adversarial training and skip connections.
`Skip', `Adv' and `GCAA' refer to the skip connections, adversarial training and global context-aware attention techniques respectively ({\em c.f.}, Section~\ref{sec:method}). The numbers prefixed with `Pretrain ' refer to the three different pretraining variants described in Section~\ref{sec:pretraining}. We observe that our proposed techniques give significant improvements over the state-of-the-art HAN network on \emph{all five} metrics. Pretrain (multi) consistently performs the best across most of the event-based metrics. (We hypothesize that using AVG pretraining objectives at the granularity of audio and video snippets helped benefit both event and segment-level performance.) Inclusion of `Adv+Skip+GCAA' improves segment-level and event-level F1 scores on three of five evaluation metrics. 
\vspace{-5mm}
\paragraph*{Exploring heterogeneous clues for weakly-supervised  audio-visual  video  parsing \cite{wu2021explore}: }
This is a recent and parallel work on AVVP.  In this paper, they tackle the issue of unreliable labels by leveraging the cross-modal correspondence of audio and visual signals.  They generate reliable event labels for each modality by swapping audio and visual tracks with other unrelated videos. Interestingly, while label cleaning helped visual modality significantly, the gains wrt to audio modality are relatively lesser.

Interestingly, while pretrained features help significantly with visual events, they do not improve performance on audio events. To understand this phenomenon better, we show category-wise F1 scores in Figure~\ref{fig:av_res} for both audio and visual modalities. As per Figure~\ref{fig:stats}, 
77\% of the `speech' category events exist  in the audio-only modality, 
possibly therefore `speech' is not helped by pretraining on  audio and visual events. Further, it is the most dominant (comprising more than 28\% of all events), possibly explaining the overall drop in Table~\ref{tab:avsegment}. Telephone ringing (that typically appears in the background) is an audio event that degrades most in performance with the pretrained model. Baby laughter and clapping, which are inherently rich in  audio and visual cues, benefit the most from cross-modal pretraining.
\begin{table}[h!]
\resizebox{\linewidth}{!}{
\begin{tabular}{c|ccccc}
Method/Event & Audio  & Visual & AV     & Ty@AV      & Ev@AV     \\ \hline
HAN \cite{tian2020avvp}    & 60.1   & 52.9   & 48.9   & 54.0     & 55.4 \\ 
Wu et al.\cite{wu2021explore} &  60.3   & 60.0   & 55.1   & 58.9 & 57.9 \\ \hline
Base      & 60.0 & 52.6   & 48.4   & 53.7   & 54.8   \\ 
Base + GCAA    & \textbf{62.0}   & 53.4   & 48.9   & 54.8   & \textbf{56.3}   \\ 
Base + Adv+Skip & 61.5   & 53.3   & 48.8 & 54.5 & 56   \\
Base + Adv+Skip+GCAA & 61.6   & \textbf{54.7}   & \textbf{50.3} & \textbf{55.5} & 56.1   \\ \hline
Pretrain (uni)      & 59.8   & 54.7     & 49.4   & 54.7   & 56.2   \\ 
Pretrain (cross)      & 60.3   & \textbf{55.1}     & 49.9  & 55.1   & \textbf{56.4}   \\ 
Pretrain (multi)      & \textbf{60.8}  & 54.8    & \textbf{50 } & \textbf{55.2}   & 56.2 \\ \hline

\end{tabular}
}
\caption{AVVP segment-level results on the LLP test set.}
\label{tab:avsegment}
\end{table}

\begin{table}[h!]
\resizebox{\linewidth}{!}{
\begin{tabular}{c|ccccc}
Method/Event & Audio  & Visual & AV     & Ty@AV      & Ev@AV     \\ \hline
HAN \cite{tian2020avvp} & 51.3   & 48.9   & 43.0   & 47.7   & 48.0     \\
Wu et al.\cite{wu2021explore} & 53.6   & 56.4   & 49.0   &  53.0 & 50.6  \\
\hline
Base & 51.9   & 48.5   & 41.9     & 47.4   & 49.5   \\
Base + GCAA & 53.4   & 48.9   & 42.0   & 48.1   & 50.7   \\
Base + Adv+Skip & 53.4   & 48.8   & 42.0 & 48.0 & 50.5   \\
Base + Adv+Skip+GCAA & \textbf{53.7}   & \textbf{50.5} & \textbf{43.5}   & \textbf{49.2} & \textbf{50.9} \\ 
\hline
Pretrain (uni)       & 50.8   & 51.5   & 42.9   & 48.4   & 49.3   \\ 
Pretrain (cross)      & 52.3   &   \textbf{52.1 }  &  43.9  & 49.4  &  50.4  \\ 
Pretrain (multi)      &  \textbf{53.4} &  51.6    & \textbf{44.3}   &  \textbf{49.8} &  \textbf{51.2}  \\ \hline
\end{tabular}
}
\caption{AVVP event-level results on the LLP test set.}
\label{tab:avevent}

\end{table}

\section{Conclusions}
In this paper, we present improved techniques for a relatively new task of Audio-Visual Video Parsing. 
We address issues in existing cross-modal feature learning that might benefit the visual streams at the cost of event detection performance in audio modality. 
We demonstrate significant improvements over the state-of-the-art for AVVP and present comprehensive ablations.

\section{Acknowledgements \label{sec:ack}}
We thank anonymous reviewers for providing constructive feedback. Preethi and Ganesh are grateful to IBM Research, India (specifically the IBM AI Horizon Networks - IIT Bombay initiative) for their support and sponsorship.

\bibliographystyle{IEEEtran}
\bibliography{mybib}

\begin{thebibliography}{10}
\providecommand{\url}[1]{#1}
\csname url@samestyle\endcsname
\providecommand{\newblock}{\relax}
\providecommand{\bibinfo}[2]{#2}
\providecommand{\BIBentrySTDinterwordspacing}{\spaceskip=0pt\relax}
\providecommand{\BIBentryALTinterwordstretchfactor}{4}
\providecommand{\BIBentryALTinterwordspacing}{\spaceskip=\fontdimen2\font plus
\BIBentryALTinterwordstretchfactor\fontdimen3\font minus
  \fontdimen4\font\relax}
\providecommand{\BIBforeignlanguage}[2]{{%
\expandafter\ifx\csname l@#1\endcsname\relax
\typeout{** WARNING: IEEEtran.bst: No hyphenation pattern has been}%
\typeout{** loaded for the language `#1'. Using the pattern for}%
\typeout{** the default language instead.}%
\else
\language=\csname l@#1\endcsname
\fi
#2}}
\providecommand{\BIBdecl}{\relax}
\BIBdecl

\bibitem{tian2020avvp}
Y.~Tian, D.~Li, and C.~Xu, ``Unified multisensory perception: Weakly-supervised
  audio-visual video parsing,'' in \emph{ECCV}, 2020.

\bibitem{Ephrat_2018}
A.~Ephrat, I.~Mosseri, O.~Lang, T.~Dekel, K.~Wilson, A.~Hassidim, W.~T.
  Freeman, and M.~Rubinstein, ``Looking to listen at the cocktail party,''
  \emph{ACM Transactions on Graphics}, vol.~37, no.~4, p. 1–11, Aug 2018.

\bibitem{vgg}
S.~{Liu} and W.~{Deng}, ``Very deep convolutional neural network based image
  classification using small training sample size,'' in \emph{Asian Conference
  on Pattern Recognition (ACPR)}, 2015, pp. 730--734.

\bibitem{gcaa}
H.~Xuan, Z.~Zhang, S.~Chen, J.~Yang, and Y.~Yan, ``Cross-modal attention
  network for temporal inconsistent audio-visual event localization,'' in
  \emph{Proceedings of the AAAI Conference on Artificial Intelligence},
  vol.~34, no.~01, 2020, pp. 279--286.

\bibitem{1039875}
M.~R. {Naphade} and T.~S. {Huang}, ``Discovering recurrent events in video
  using unsupervised methods,'' in \emph{Proceedings. International Conference
  on Image Processing}, vol.~2, 2002, pp. II--II.

\bibitem{NEURIPS2018_c4616f5a}
B.~Korbar, D.~Tran, and L.~Torresani, ``Cooperative learning of audio and video
  models from self-supervised synchronization,'' in \emph{Advances in Neural
  Information Processing Systems}, 2018.

\bibitem{1a18f054b93d4818af72164f56836616}
J.~Vroomen, M.~Keetels, B.~{de Gelder}, and P.~Bertelson,
  ``\BIBforeignlanguage{English}{Recalibration of temporal order perception by
  exposure to audio-visual asynchrony},''
  \emph{\BIBforeignlanguage{English}{Cognitive Brain Research}}, 2004.

\bibitem{Xuan_Zhang_Chen_Yang_Yan_2020}
H.~Xuan, Z.~Zhang, S.~Chen, J.~Yang, and Y.~Yan, ``Cross-modal attention
  network for temporal inconsistent audio-visual event localization,''
  \emph{AAAI}, 2020.

\bibitem{dcma}
Y.-H. Lee, D.-W. Jang, J.-B. Kim, R.-H. Park, and H.-M. Park, ``Audio–visual
  speech recognition based on dual cross-modality attentions with the
  transformer model,'' \emph{Applied Sciences}, vol.~10, no.~20, 2020.

\bibitem{mil}
O.~Maron and T.~Lozano-P{\'e}rez, ``A framework for multiple-instance
  learning,'' \emph{Advances in neural information processing systems}, pp.
  570--576, 1998.

\bibitem{tian2018audiovisual}
Y.~Tian, J.~Shi, B.~Li, Z.~Duan, and C.~Xu, ``Audio-visual event localization
  in unconstrained videos,'' in \emph{ECCV}, 2018.

\bibitem{audiovisualscene}
A.~Owens and A.~A. Efros, ``Audio-visual scene analysis with self-supervised
  multisensory features,'' in \emph{Proceedings of the European Conference on
  Computer Vision (ECCV)}, September 2018.

\bibitem{avid}
P.~Morgado, N.~Vasconcelos, and I.~Misra, ``Audio-visual instance
  discrimination with cross-modal agreement,'' 2020.

\bibitem{avcluster}
H.~Alwassel, D.~Mahajan, B.~Korbar, L.~Torresani, B.~Ghanem, and D.~Tran,
  ``Self-supervised learning by cross-modal audio-video clustering,'' in
  \emph{Advances in Neural Information Processing Systems (NeurIPS)}, 2020.

\bibitem{resnet152}
K.~He, X.~Zhang, S.~Ren, and J.~Sun, ``Deep residual learning for image
  recognition,'' in \emph{2016 IEEE Conference on Computer Vision and Pattern
  Recognition (CVPR)}, 2016, pp. 770--778.

\bibitem{r2p1d}
R.~Hou, C.~Chen, R.~Sukthankar, and M.~Shah, ``An efficient 3d {CNN} for
  action/object segmentation in video,'' in \emph{30th British Machine Vision
  Conference 2019, {BMVC} 2019, Cardiff, UK, September 9-12, 2019}.\hskip 1em
  plus 0.5em minus 0.4em\relax {BMVA} Press, 2019, p. 170.

\bibitem{gradientReversal16}
Y.~Ganin, E.~Ustinova, H.~Ajakan, P.~Germain, H.~Larochelle, F.~Laviolette,
  M.~Marchand, and V.~S. Lempitsky, ``Domain-adversarial training of neural
  networks,'' \emph{JMLR}, vol.~17, 2016.

\bibitem{uniter}
Y.-C. Chen, L.~Li, L.~Yu, A.~E. Kholy, F.~Ahmed, Z.~Gan, Y.~Cheng, and J.~Liu,
  ``Uniter: Universal image-text representation learning,'' in \emph{ECCV},
  2020.

\bibitem{audioset}
J.~F. Gemmeke, D.~P.~W. Ellis, D.~Freedman, A.~Jansen, W.~Lawrence, R.~C.
  Moore, M.~Plakal, and M.~Ritter, ``Audio set: An ontology and human-labeled
  dataset for audio events,'' in \emph{Proc. IEEE ICASSP 2017}, New Orleans,
  LA, 2017.

\bibitem{wu2021explore}
Y.~Wu and Y.~Yang, ``Exploring heterogeneous clues for weakly-supervised
  audio-visual video parsing,'' in \emph{Proceedings of the IEEE Conference on
  Computer Vision and Pattern Recognition (CVPR)}, 2021.

\end{thebibliography}

\end{document}